\documentclass[sigconf,screen]{acmart}
\acmBooktitle{Companion Proceedings of the 34th ACM Symposium on the Foundations of Software Engineering (FSE '26), June 5--9, 2026, Montreal, Canada}
\AtBeginDocument{%
  }
    
\setcopyright{acmlicensed}
\copyrightyear{2026}
\acmYear{2026}
\acmDOI{10.1145/3803437.3805583} 


\usepackage{balance}

\begin{document}

\copyrightyear{2026}
\acmYear{2026}
\setcopyright{cc}
\setcctype{by}
\acmConference[FSE Companion '26]{34th ACM Joint European Software Engineering Conference and Symposium on the Foundations of Software Engineering}{July 05--09, 2026}{Montreal, QC, Canada}
\acmBooktitle{34th ACM Joint European Software Engineering Conference and Symposium on the Foundations of Software Engineering (FSE Companion '26), July 05--09, 2026, Montreal, QC, Canada}
\acmDOI{10.1145/3803437.3805583}
\acmISBN{979-8-4007-2636-1/2026/07}

%%
%% The "title" command has an optional parameter,
%% allowing the author to define a "short title" to be used in page headers.
%\title{}

\title[Towards Predicting Multi-Vulnerability Attack Chains]{Towards Predicting Multi-Vulnerability Attack Chains in Software Supply Chains from Software Bill of Materials Graphs}

%%
%% The "author" command and its associated commands are used to define
%% the authors and their affiliations.
%% Of note is the shared affiliation of the first two authors, and the
%% "authornote" and "authornotemark" commands
%% used to denote shared contribution to the research.
\author{Laura Baird}
\email{lbaird@uccs.edu}
\orcid{0009-0002-0538-7366}
\affiliation{
  \institution{University of Colorado Colorado Springs (UCCS)}
  \city{Colorado Springs}
  \state{Colorado}
  \country{USA}
}

\author{Armin Moin}
\email{amoin@uccs.edu}
\orcid{0000-0002-8484-7836}
\affiliation{
  \institution{University of Colorado Colorado Springs (UCCS)}
  \city{Colorado Springs}
  \state{Colorado}
  \country{USA}
}

%%
%% By default, the full list of authors will be used in the page
%% headers. Often, this list is too long, and will overlap
%% other information printed in the page headers. This command allows
%% the author to define a more concise list
%% of authors' names for this purpose.
\renewcommand{\shortauthors}{Baird and Moin}

%%
%% The abstract is a short summary of the work to be presented in the
%% article.
\begin{abstract}
Software supply chain security compromises often stem from cascaded interactions of vulnerabilities, for example, between multiple vulnerable components. Yet, Software Bill of Materials (SBOM)-based pipelines for security analysis typically treat scanner findings as independent per-CVE (Common Vulnerabilities and Exposures) records. We propose a new research direction based on \textit{learning} multi-vulnerability attack chains through a novel SBOM-driven graph-learning approach. This treats SBOM structure and scanner outputs as a dependency-constrained evidence graph rather than a flat list of vulnerabilities. We represent vulnerability-enriched CycloneDX SBOMs as heterogeneous graphs whose nodes capture software components and known vulnerabilities (i.e, CVEs), connected by typed relations, such as dependency and vulnerability links. We train a Heterogeneous Graph Attention Network (HGAT) to predict whether a component is associated with at least one known vulnerability as a feasibility check for learning over this structure. Additionally, we frame the discovery of cascading vulnerabilities as CVE-pair link prediction using a lightweight Multi-Layer Perceptron (MLP) neural network trained on documented multi-vulnerability chains. Validated on 200 real-world SBOMs from the Wild SBOMs public dataset, the HGAT component classifier achieves 91.03\% Accuracy and 74.02\% F1-score, while the cascade predictor model (MLP) achieves a Receiver Operating Characteristic - Area Under Curve (ROC-AUC) of 0.93 on a seed set of 35 documented attack chains.
\end{abstract} 


\maketitle

\section{Introduction} \label{sec:introduction}
Software Bills of Materials (SBOMs) are increasingly used to support supply chain transparency and security management. For instance, the U.S. federal government has previously mandated, and currently may mandate or encourage SBOM delivery from its software system contractors, depending on risk assessment \cite{TheWhiteHouse2021, TheWhiteHouse2026}. However, SBOM generation and maintenance remain difficult in realistic environments~\cite{Xia+2023,Stalnaker+2024}, and tool choices can materially affect downstream vulnerability detection outcomes~\cite{ODonoghue+2024,Benedetti+2024}. As a result, the practical \textit{SBOM-in, vulnerabilities-out} pipeline can be noisy and variable even before any higher-order reasoning is attempted. These issues motivate a representation that treats SBOM structure and scanner outputs as evidence that must be interpreted in a supply chain dependency context, rather than as isolated and independent records.

However, the status quo is different: Most current software supply chain security analysis practices and tools still focus on individual entries of the Common Vulnerabilities and Exposures (CVE)~\cite{MITRE2025a} database in isolation, as well as on per-CVE scoring, such as CVSS, EPSS, and VEX-style annotations~\cite{FIRST2024,NTIA2021,FIRST2021}. This is useful for vulnerability triage, but it is fundamentally misaligned with a recurring failure mode: \textit{cascaded} attacks where multiple vulnerabilities (often across different components) interact to enable an exploitation path. Static scanners such as Snyk and Trivy, primarily report per-component CVE lists and therefore do not surface cross-component chains~\cite{Snyk2025,Aquasecurity2025}. Consequently, issues that look low-severity in isolation can become critical when composed through the dependency structure and security weakness patterns (e.g., from the Common Weakness Enumeration, CWE, database \cite{MITRE2025}).

Real-world incidents show how chaining amplifies risk beyond what per-CVE scores imply. For example, the 2021 ProxyLogon campaign chained four Microsoft Exchange vulnerabilities (CVE-2021-26855/-26857/-26858/-27065) into a pre-authentication path to persistent remote control of enterprise email servers~\cite{Microsoft2021}. More broadly, vulnerable dependencies are often transitive and can persist for long periods, which increases the practical likelihood of cascades across dependency graphs~\cite{Kumar+2024}. This motivates shifting from a ``score each CVE'' philosophy to ``learn interaction patterns constrained by the dependency graph.''

We propose a graph-learning approach that models an SBOM as a \textit{heterogeneous graph} over components, vulnerabilities, and weaknesses. We deploy a Heterogeneous Graph Attention Network (HGAT)~\cite{Yang+2021} model to learn representations over this structure and perform component-level feasibility checks. Additionally, to make progress on cascaded chain discovery while the available data for training is limited, we frame cascade discovery as a link prediction problem over CVE pairs using a lightweight feature-based Multi-Layer Perceptron (MLP) neural network model. The ranked CVE links can then be composed into candidate multi-step chains for analyst triage and for future benchmarking.

\begin{figure}[t]
  \centering
  \includegraphics[width=0.82\linewidth]{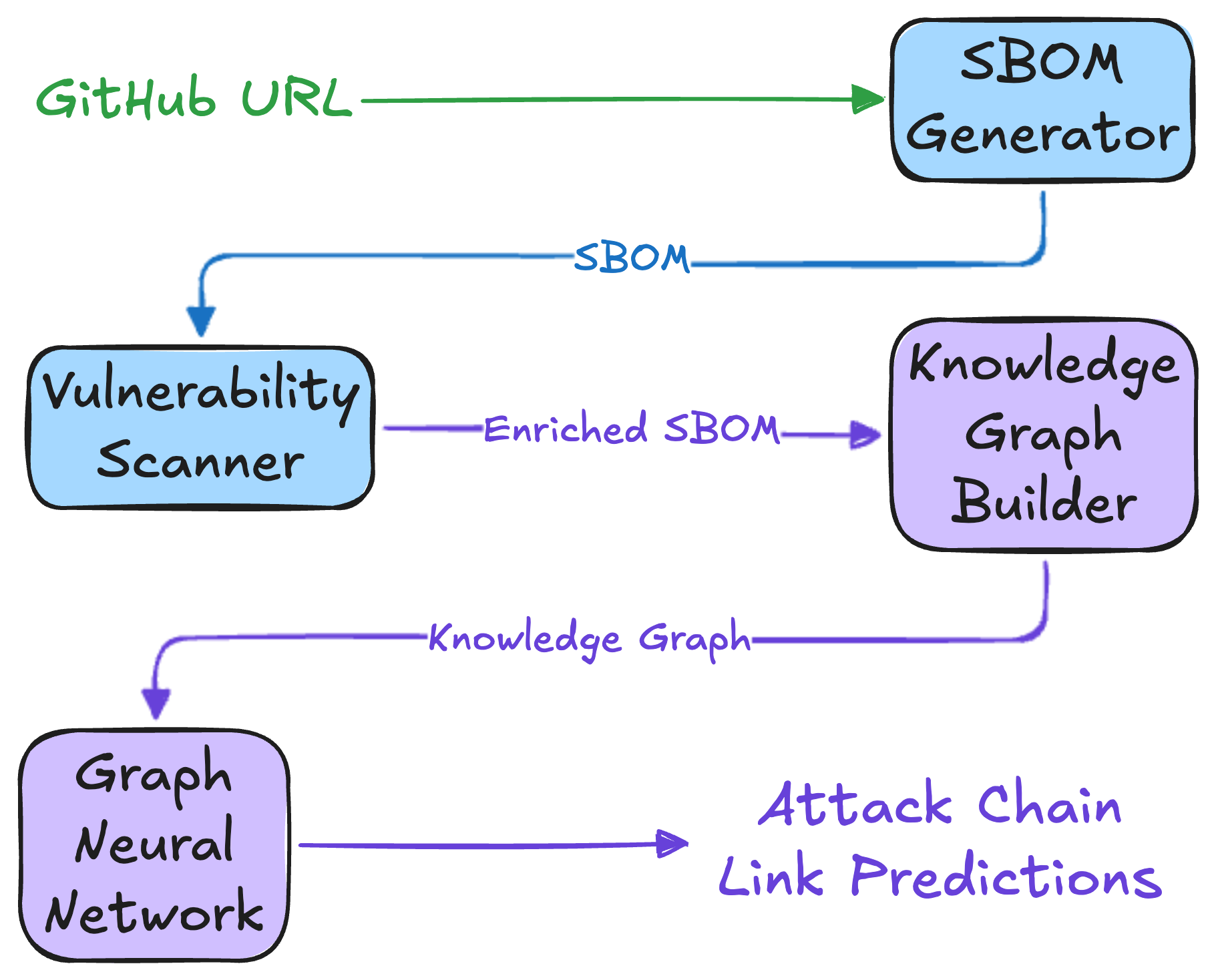}
  \caption{The architecture of the pipeline: from a GitHub repository to attack chain predictions.}
  \Description{Flow diagram from GitHub URL to SBOM Generator and Vulnerability Scanner producing a vulnerability-enriched SBOM; into a KG Builder that outputs a KG to a Graph Neural Network to generate attack chain link predictions.}
  \label{fig:sbom-toolkit}
\end{figure}

Our current prototype decomposes this goal into two modular learning stages. First, an HGAT learns representations over the heterogeneous SBOM graph and serves as a feasibility check that the dependency structure carries a signal beyond local component metadata. Second, a lightweight CVE-pair predictor independently learns a co-exploitation prior from scarce documented chains, using metadata features from the records of the National Vulnerability Dataset (NVD)~\cite{NIST2024}, which is very similar to the CVE database, rather than HGAT embeddings, and produces ranked candidate links. However, in the intended end-to-end workflow, HGAT-learned embeddings would feed directly into the cascade predictor, and predicted CVE-pair links would be projected back onto SBOM dependency subgraphs to prioritize component-level cascade candidates for analyst triage.

In proposing the stated new research direction, this paper makes the following key contributions: (i) It proposes a pipeline that converts vulnerability-enriched CycloneDX-formatted SBOMs into heterogeneous graphs with typed nodes and edges. This can be independently used for various research purposes in this field. (ii) It conducts a feasibility study on the heterogeneous graph with real SBOMs, including a targeted ablation study to test whether the dependency structure carries a useful signal. Then, an MLP-based CVE-pair cascade predictor that yields ranked co-exploitation candidates for vulnerabilities and is evaluated on a seed set of documented multi-CVE chains makes the final attack chain link predictions. (iii) It also proposes an analyst-facing inspection workflow (mapping protocol) that localizes the documented attack chain evidence to SBOM subgraphs and supports lightweight labeling, error analysis, and top-$k$ evaluation.

The remainder of this paper is structured as follows. Section~\ref{sec:related-work} reviews related work. Section~\ref{sec:approach} presents the new research direction and describes the evidence-graph construction and learning pipeline. Further, Section~\ref{sec:experimental-results} reports some preliminary findings. Finally, Section~\ref{sec:conclusion-future-work} concludes and outlines the future work.

\section{Related Work} \label{sec:related-work}

\paragraph{SBOM quality and variability in vulnerability pipelines}
SBOM adoption has accelerated, but empirical work highlights persistent challenges in generating, validating, and maintaining SBOMs at scale~\cite{Xia+2023,Stalnaker+2024}. Multiple studies report that SBOM generator and analysis tool choices can change downstream vulnerability findings, increasing both false positives and false negatives depending on ecosystem and configuration~\cite{ODonoghue+2024,Benedetti+2024}. Our work is compatible with this reality: rather than treating scanner outputs as independent records, we model the dependency-constrained relationships among components and security metadata, which can support robustness checks and downstream reasoning about cascades.

\paragraph{Vulnerability detection and prioritization beyond per-CVE scores}
Industrial practice still revolves around CVE-level reporting and prioritization mechanisms (e.g., CVSS, EPSS, and VEX-style exploitability/exposure annotations)~\cite{FIRST2024,NTIA2021,FIRST2021}. These approaches provided useful results, but they did not explicitly model cross-component composition. Dependency scanners (e.g., Snyk, Trivy) are effective at enumerating vulnerable components but generally do not infer multi-step chains that emerge from graph structure~\cite{Snyk2025,Aquasecurity2025}. Recent research agendas in supply chain security called for reasoning across components, ecosystems, and attacker techniques rather than treating severity scores as sufficient~\cite{Williams+2025}.

\paragraph{Graph learning for vulnerability reasoning}
Graph Neural Networks (GNNs) have recently shown promise for representing security entities and learning interaction patterns. Yin et al.~\cite{Yin+2023a,Yin+2023} proposed heterogeneous-graph-driven exploitability prediction and co-exploitation modeling as link prediction over vulnerability Knowledge Graphs (KGs). Their results supported the broader claim that relational structures improve inference relative to isolated records. Our work differs in that it situates reasoning within SBOM dependency constraints: We model components, vulnerabilities, and weaknesses jointly, and we target cascaded behavior that can span multiple components.

% \paragraph{LLMs, grounding, and tool-based interfaces}
% Large Language Models (LLMs) can support analyst workflows when grounded against authoritative data sources. KG-enhanced retrieval has been used to improve attribution and reconstruction tasks~\cite{Kurniawan+2024}. Separately, tool-calling approaches (e.g., ReAct or Toolformer) illustrated how LLMs can interface with structured systems~\cite{Yao+2023,Schick+2023}. In this work, we treat LLM-backed tool-based inspection as \textit{future workflow support}: a way to explain and audit predicted candidates against SBOM/KG facts, rather than as a core evaluated contribution in the current prototype.

\paragraph{Supply chain vulnerability analysis tooling}
Tools such as Dependency Track and OSS Review Toolkit visualize vulnerable dependencies and CVE propagation across supply chains~\cite{OWASP2025a,ORT2025}. These are valuable for triage, but they rely on static heuristics rather than learned models of vulnerability interactions. Separately, LLM-backed tool-calling workflows (e.g., ReAct~\cite{Yao+2023}) offer a complementary direction for grounding inspection against structured data~\cite{Kurniawan+2024,Schick+2023}. Extending our proposed approach in this direction is among the future work mentioned in Section \ref{sec:conclusion-future-work}).

\paragraph{Research data for cascaded vulnerability evaluation}
A key barrier in this field is the scarcity of publicly documented, semantically annotated multi-vulnerability chains. VulKG \cite{Yin+2024} provided large-scale vulnerability graphs that were useful for relational learning, and Wild SBOMs \cite{Soeiro+2025a} provided large-scale SBOM data for realistic graph construction. For incident-level chain documentation, we plan to draw on curated public corpora, such as the Atlantic Council supply chain attack dataset~\cite{Messieh2023}. Our current evaluation uses a small seed set of multi-CVE chains extracted from disclosures and incident reports as documented in our open dataset \cite{DVN/A6CZRB_2026}.

\section{Proposed Research Direction} \label{sec:approach}
Our proposed novel research direction involves shifting supply chain security analysis from per-CVE scoring and analysis to \textit{dependency constrained relational modeling} that can surface candidate attack cascades. Figure~\ref{fig:sbom-toolkit} presents the overall architecture of the pipeline. Here are the four key steps in this process: i) We construct the enriched-SBOM evidence graph, which includes the CVE/CWE data for each software component (dependency) in the SBOM. ii) We \textit{learn} the vulnerability of software components in the SBOM through an HGAT (i.e., classify each component into vulnerable or safe classes). iii) We predict vulnerability link pairs and rank candidate cascade links to prioritize the highest-risk chains using an MLP. iv) Finally, we project those candidates onto SBOM dependency structures for manual or automated inspection. 

In the following, we elaborate on these steps.

\paragraph{Step I}

\begin{figure}[t]
  \centering
  \includegraphics[width=0.80\linewidth]{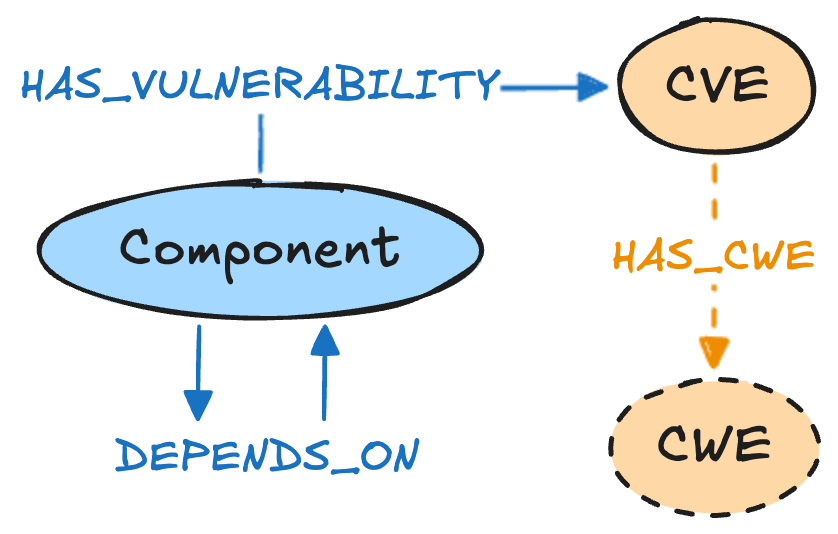}
  \caption{The basic schema of the proposed knowledge graph.}
  \Description{Schema diagram showing Component, CVE, and CWE node types connected by DEPENDS\_ON, HAS\_VULNERABILITY, and HAS\_CWE edges. CWE and HAS\_CWE are shown with dashed lines indicating planned support.}
  \label{fig:kg-base}
\end{figure}

Given a repository URL, we generate an SBOM in CycloneDX format using Syft~\cite{Anchore2025a}, and enrich components with vulnerability information using Grype~\cite{Anchore2025}. Grype is backed by the Open Source Vulnerabilities (OSV) database~\cite{Google2025}, allowing us to associate components (e.g., package name and version) with known CVEs. We convert each enriched SBOM into a heterogeneous graph (also representable as a KG). Figure~\ref{fig:kg-base} shows the base schema. The node types are as follows: (i) \textbf{Component} nodes, which are software packages or SBOM dependencies, (ii) \textbf{CVE} nodes (i.e., known vulnerabilities documented in the CVE entries of the OSV database), and (iii) \textbf{CWE} nodes (i.e., identified weakness types in the CWE database). The dashed elements in Figure~\ref{fig:kg-base}, that is, CWE nodes and \texttt{HAS\_CWE} edges, are supported by the schema but are not yet implemented in the current prototype. Moreover, we include typed edges as follows: \textbf{DEPENDS\_ON} between components, based on the SBOM dependency relations, \textbf{HAS\_VULNERABILITY} from components to CVEs, according to the scanner, and \textbf{HAS\_CWE} from CVEs to CWEs, relying on the CWE information that exists in the CVE entries. In the present prototype, the enriched SBOMs predate a CWE extraction step added to our scanning pipeline. Consequently, \textit{HAS\_CWE} edges are absent from the current graphs. Re-scanning with CWE extraction enabled, or resolving CWE mappings from CVE records during graph construction, would populate this data and is planned for extended evaluation. Note that we use lightweight, interpretable features derived from the SBOM data and from the vulnerability annotations in the enriched SBOM: \textit{Component nodes} include summarized vulnerability metadata (e.g., CVSS aggregates and severity indicators), dependency metadata (e.g., direct vs. transitive dependency), graph node degree statistics, and software license indicators; \textit{CVE nodes} encode vulnerability severity score and temporal metadata (e.g., CVSS-derived bins and recency when available); \textit{CWE nodes} use simple graph features (e.g., normalized degree/count of mapped CVEs within the SBOM graph) when CWE mappings are available in the scanner output.

\paragraph{Step II}

We deploy an HGAT \cite{Yang+2021} model to learn the vulnerability status of the components based on the heterogeneous graph structure. The \textit{attention} mechanism in HGAT enables relation-aware neighbor aggregation, allowing the model to weight different edge types (e.g., component dependencies vs. vulnerability links) differently. Our hyperparameters are as follows: A 2-layer HGAT with 2 multi-head attention~\cite{Velickovic+2018} heads, hidden dimension 64, dropout 0.2, trained with the Adam optimizer (lr=$10^{-3}$, weight decay $5{\times}10^{-4}$) and cross-entropy loss for up to 30 epochs on batches of 2 graphs. As a feasibility check, we predict a coarse component label: \texttt{has-any-CVE} (whether a component is associated with at least one known CVE in the enriched SBOM). This task is not intended to discover previously unidentified component vulnerabilities. Instead, we use it as a mechanism to conduct our ablation study to demonstrate the benefits of including the dependency information. As expected, the classification performance of the HGAT model significantly degrades when dependency information is missing.

\paragraph{Step III}

Directly learning multi-step cascaded attack chains is difficult because fully documented real-world chains are scarce and rarely mapped to precise dependency contexts. To make progress with limited data, we model cascade discovery as \textit{link prediction over CVE pairs}. Given two CVEs, the task is to estimate whether they are plausibly co-exploited as part of a chained exploitation scenario. High-scoring links can then be composed into multi-step candidate chains by chaining edges. We extract CVE sequences from public advisories and incident reports that explicitly describe chained exploitation. Our seed set contains 35 chains (27 from vulnerability disclosures, and 8 from incident reports), with chain lengths ranging from 2 to 4 CVEs (median 2, and mean 2.51). The curated chain corpus and the feature extraction scripts are available in our open dataset~\cite{DVN/A6CZRB_2026} and GitHub repository~\cite{QASLab2026}, respectively. Each documented chain yields within-chain CVE pairs as positive examples. We construct negative examples by sampling CVE pairs not observed within the same chain at a 2:1 negatives-to-positives ratio. We chose 2:1 as a moderate compromise between a balanced setting (which underrepresents the real-world base rate of non-chain pairs) and aggressive oversampling (which risks overwhelming the scarce positive signal at this dataset scale). This ratio is parameterized in our pipeline. We plan to perform sensitivity analysis across ratios in the future. We implement a lightweight feature-based MLP neural network model that consumes a 22-dimensional feature vector per CVE pair: 9 per-CVE features (i.e., CVSS score, severity one-hot, year, exploited flag, reference count, and CWE count) derived from the NVD~\cite{NIST2024} metadata, plus 4 interaction features (CVSS difference, CVSS product, year gap, and both-exploited indicator). The MLP architecture consists of 5 layers with 22, 64, 32, 16, and 1 neurons, respectively. Also, the hyperparameters are as follows: dropout 0.3, trained with Adam optimizer (lr=$10^{-3}$), and binary cross-entropy loss for up to 50 epochs with early stopping (patience 10 on validation ROC-AUC). The output is a co-exploitation probability score used to rank candidate links. This enables predicting co-exploitation for pairs of vulnerabilities, as well as multi-step attack chains. Our current prototype implementation, which is a work in progress, has achieved the former. The latter remains as future work (see Section \ref{sec:conclusion-future-work}).

\paragraph{Step IV}

Given a documented CVE chain from an advisory or incident report and an SBOM, we determine which components in the SBOM carry the chain's CVEs and how those components are connected in the dependency graph. We intersect the CVEs with scanner-produced \texttt{HAS\_VULNERABILITY} edges in the SBOM to identify the specified versioned components carrying each CVE. We then project the mapped components onto the \texttt{DEPENDS\_ON} graph and extract the induced subgraph spanning those components. We treat the resulting subgraph as the SBOM-localized context for the chain.

\section{Preliminary Results} \label{sec:experimental-results}

Our evaluation focuses on two Research Questions (RQs): (RQ1) Does the heterogeneous SBOM graph representation support basic learning on real-world SBOMs, and does dependency structure matter under ablation? (RQ2) Can we obtain a signal for cascade discovery via CVE-pair prediction from a small set of documented chains? 

%These stages are currently evaluated independently: Stage~1 tests whether the SBOM evidence graph supports learning over dependency structure, and Stage~2 tests whether scarce chain evidence can train a useful co-exploitation prior. Connecting the two---using HGAT embeddings as cascade predictor features and projecting cascade predictions onto SBOM graphs---is planned future work.

%\paragraph{HGAT feasibility on real SBOM graphs}

We use the publicly available \textit{Wild SBOMs} dataset~\cite{Soeiro+2025a} for the HGAT evaluation. We filter to CycloneDX-format SBOMs for Python projects and use a subset of 200 randomly selected SBOMs for our current experiments. The filtered subset, train/val/test splits, and trained model checkpoints are provided in our dataset~\cite{DVN/A6CZRB_2026}. We chose CycloneDX because it is the format most represented in the Wild SBOMs corpus for Python; the downstream graph schema is not format-specific, and adapting the ingestion parser for SPDX SBOMs would be straightforward. We split SBOMs into training, validation, and testing sets (70\%, 15\%, and 15\%, respectively). The HGAT achieves 91.03\% Accuracy, 80.84\% Precision, 68.26\% Recall, and 74.02\% F1-score on the \texttt{has-any-CVE} component label. To test whether dependency structure contributes a signal beyond local metadata, we ablate \texttt{DEPENDS\_ON} edges by masking them at inference time on the trained model (i.e., zeroing the edge index without retraining). This ablation sharply reduces positive predictions (Recall collapses to 0), suggesting the model relies on relational structure rather than only node-local signals. Importantly, the resulting high Accuracy under ablation is consistent with class imbalance (predicting the majority class), so Recall and F1 are the more informative indicators for this test.

\begin{table}[t]
  \centering
  \setlength{\tabcolsep}{4pt}
  \caption{Component classification metrics on 200 Python CycloneDX SBOMs from Wild SBOMs~\cite{Soeiro+2025a}.}
  \label{tab:node-classification}
  \begin{tabular}{lcccc}
  \hline
  Setting & Accuracy & Precision & Recall & F1 \\
  \hline
  HGAT (full graph) & 0.9103 & 0.8084 & 0.6826 & 0.7402 \\
  No dependency edges & 0.8128 & -- & 0.0000 & -- \\
  \hline
  \end{tabular}
\end{table}

%\paragraph{CVE-pair cascade predictor on documented chains}
We also evaluate the MLP-based CVE-pair cascade predictor on a seed set of 35 documented multi-CVE chains described in Section~\ref{sec:approach}, paired with sampled non-chain CVE pairs (2:1 negatives to positives). On this initial setting, the model achieves 0.93 ROC-AUC. We acknowledge that the ROC-AUC metric may achieve high performance but be misleading in the case of unbalanced classes in the data. Moreover, our preliminary results indicate the model's ability to distinguish metadata patterns of co-exploited CVE pairs from random pairs. However, it does not yet demonstrate the prediction of novel, previously undocumented chains. This evaluation is limited by chain scarcity and by pair-level splitting: the same CVE may appear in both train and test via different pairings, which can inflate apparent generalization. We implement chain-level and temporal splits to better measure generalization to unseen cascades.

\section{Conclusion and Future Work} \label{sec:conclusion-future-work}
We have proposed a new research direction in software supply chain security that emphasizes complex interactions among vulnerabilities and security weaknesses that span multiple components in the supply chain. Our proposed novel approach models the SBOM as a heterogeneous graph and operationalizes cascading vulnerability discovery as CVE-pair link prediction. Our preliminary results demonstrated that the software dependency structure carries a learnable signal. This was achieved using the HGAT model. Also, it demonstrated that even scarce chain documentation can train a co-exploitation prior with promising discriminative power. This was achieved using the MLP model. 

We note that the two models currently operate on disjoint CVE populations and are not yet connected end-to-end. Bridging this gap by extending the MLP link predictor from pair-level co-exploitation scoring to the intended multi-step chain assembly by composing high-scoring pairs into end-to-end attack chains will be our key future work. Other future works include expanding the documented chain corpus, implementing chain-level and temporal splits and sensitivity analysis across negative sampling ratios to better assess generalization and robustness, extending ingestion to SPDX-format SBOMs and ecosystems beyond Python to evaluate generalization, re-scanning SBOMs with CWE extraction enabled for investigating CWE-type co-occurrence patterns within documented chains to improve interpretability, and evaluating LLM-based baselines (e.g., ReAct~\cite{Yao+2023} prompting over SBOM/KG context) as a complementary approach.

\section*{Data and Source Code Availability}
All source code and data are publicly available in our GitHub repository \cite{QASLab2026} and on DataVerse \cite{DVN/A6CZRB_2026}.

\section*{Acknowledgments}
This material is based upon work supported by the U.S. National Science Foundation (NSF) under Grant No. 2349452. Any opinions, findings, conclusions, or recommendations expressed in this material are those of the authors and do not necessarily reflect the views of the NSF. Additionally, generative AI tools (OpenAI/Anthropic) were used to assist with the content and programming tasks.

\clearpage
\newpage

\balance
\bibliographystyle{ACM-Reference-Format}
\bibliography{References}

\end{document}